\documentclass[useAMS,usenatbib,letterpaper]{mn2e}
\usepackage{amsmath,amssymb}
\usepackage{graphicx}
\usepackage{natbib}

\def\Omm{{\Omega_m}}
\def\Ommz{{\Omega_m^{\,z}}}

\def\Omk{{\Omega_k}}
\def\Oml{{\Omega_{\Lambda}}}

\def\aap{A\&A}
\def\apj{ApJ}

\def\apjl{ApJ}
\def\mnras{MNRAS}

\def\aj{AJ}

\def\nat{Nat}

\def\apjs{ApJS}
\def\pasj{PASJ}
\newcommand{\beq}{
\begin{equation}
}
\newcommand{\eeq}{
\end{equation}
}
\newcommand{\kms}{\,{\rm km\,s^{-1}}}
\newcommand{\msun}{\,{\rm M_\odot}}

\def\simlt{\mathrel{\rlap{\lower 3pt\hbox{$\sim$}}\raise 2.0pt\hbox{$<$}}}
\def\simgt{\mathrel{\rlap{\lower 3pt\hbox{$\sim$}} \raise 2.0pt\hbox{$>$}}}

\title[Quasistars and the cosmic evolution of massive black holes]{Quasistars and the cosmic evolution of massive black holes}
\author[Volonteri and æBegelman]{Marta Volonteri$^{1}$\thanks{E-mail:
martav@umich.edu (MV); mitch@jila.colorado.edu (MCB)} and Mitchell C. Begelman$^{2,3}$\footnotemark[1]\\
$^{1}$Department of Astronomy, University of Michigan, 500 Church Street, Ann Arbor, MI, USA\\
$^{2}$JILA, 440 UCB, University of Colorado at Boulder, Boulder, CO 80309-0440, USA\\
$^{3}$ Department of Astrophysical and Planetary Sciences, 391 UCB, University of Colorado, Boulder, CO 80309-0391, USA}
\begin{document}

\date{}

\pagerange{\pageref{firstpage}--\pageref{lastpage}} \pubyear{2010}

\maketitle

\label{firstpage}

\begin{abstract}
We explore the cosmic evolution of massive black hole (MBH) seeds forming within `quasistars' (QSs), accreting black holes embedded within massive hydrostatic gaseous envelopes. These structures could form if the infall of gas into the center of a halo exceeds about 1 $\msun$ yr$^{-1}$. æThe collapsing gas traps its own radiation and forms a radiation pressure-supported supermassive star. When the core of the supermassive star collapses, the resulting system becomes a quasistar. We use a merger-tree approach to estimate the rate at which supermassive stars might form as a function of redshift, and the statistical properties of the resulting QS and seed black hole populations. æWe relate the triggering of runaway infall to major mergers of gas-rich galaxies, and to a threshold for global gravitational instability, which we link to the angular momentum æof the host. This is the main parameter of our models. æOnce infall is triggered, its rate is determined by the halo potential; the properties of the resulting supermassive star, QS and seed black hole depend on this rate. æAfter the epoch of QSs, we model the growth of MBHs within their hosts in a merger-driven accretion scenario. We compare MBH seeds grown inside quasistars to a seed model that derives from the remnants of the first metal-free stars, and also study the case in which both channels of MBH formation operate simultaneously. æWe find that a limited range of supermassive star/QS/MBH formation efficiencies exists that allows one to reproduce observational constraints. Our models match the density of $z=6$ quasars, the cumulative mass density accreted onto MBHs (according to So\l tan's argument), and the current mass density of MBHs. ææThe mass function of QSs peaks at $M_{\rm QS}\simeq 10^6 \msun$, and we calculate the number counts for the {\it JWST} $2-10\ \mu$m band. We find that {\it JWST} could detect up to several QSs per field at $z\simeq 5-10$. æ
\end{abstract}


\section{Introduction}

While there is ample evidence that supermassive black holes æpopulate the nuclei of most large galaxies and that some black holes with masses exceeding $10^9 \msun$ formed as early as $z \ga 6$ \citep[e.g.,][]{Fanetal2001a,Barthetal2003,Djorgovski2008,Willott2009,Jiang2009}, there is little consensus as to the progenitors of these holes. æTwo schools of thought have persisted since \cite{Rees1978} first devised a flow chart outlining possible routes of massive black hole (MBH\footnote{We refer here generically to MBHs when the hole mass is above the limit for black hole formation in today's stars, $\simeq 50 \msun$. This definition comprises both seed black holes and supermassive black holes in galaxies and quasars.}) formation. According to one line of argument, supermassive black holes grew from the remnants of an early population of massive stars, the so-called Population III (Pop III), which is believed to have formed in pregalactic minihalos at $z\ga 20$. æAccording to the other, the precursors of supermassive black holes could have formed by the `direct collapse' of large amounts of gas in much larger halos at later times. æ

Each scenario has both positive and negative attributes. æAlthough Pop III remnants were unlikely to have been more massive than a few hundred $\msun$ each, they would have formed relatively early and thus their growth process would have had a considerable head start. æThey could have congregated and merged in the cores of merging minihalos, while simultaneously growing by accretion. æModels for the growth of supermassive black holes from stellar-mass seeds \citep{MadauRees2001,VHM,Rhook2006,Monaco2007,Somerville2008,VLN2008} are moderately successful in reproducing the current-day population of supermassive black holes, but have rather more difficulty in producing enough $10^9 \msun$ holes at $z \ga 6$ to explain the earliest known quasars \citep{VR2005,Shapiro2005,VR2006}. æAdditional worries about the Pop III scenario include the possibility that too many of the remnants would have been ejected from the cores of merging halos and that their accretion rates would be depressed by the shallow potential wells of the host mini-halos and heating of the ambient gas by stellar radiation and winds \citep[and references therein]{Milos2009}. 

Seed black holes that formed by direct collapse (e.g., at $z \la 15$) would have had less time to grow, but this would have been partly compensated by their larger initial masses. æVarious direct collapse models for seed formation have been proposed \citep{LoebRasio1994, Eisenstein1995,HNR1998, BrommLoeb2003,Koushiappas2004,Begelman2006,LN2006,Begelman2010}, but so far only limited attempts have been made to place these models into the context of structure formation theories \citep[e.g.,][]{VLN2008, Lagos2008,svanwas2010}. æBecause direct collapse models draw only indirectly on star formation lore, there is much less consensus about the initial conditions for direct collapse and the details of how it might have occurred. 

In a series of recent papers, we have described a sequence of events that we believe represents a plausible route to MBH formation via direct collapse \citep{Begelman2006,Begelman2008,Begelman2010}. æIn this picture, the main triggering event is the infall of gas into the center of a halo at a rate exceeding about 1 $\msun$ yr$^{-1}$. æSuch large rates of infall are possible in halos with virial temperatures in excess of about $10^4$ K, which only become common at $z \la 10-15$. æQuestions remain about the ability of gas to accumulate at such a high rate without too much of it fragmenting into stars, but as we argue below, recent simulations as well as analytic calculations suggest that the importance of fragmentation may have been severely overestimated in past work. æThe collapsing gas traps its own radiation and forms a quasistatic, radiation pressure-supported supermassive star, which burns hydrogen for about a million years while growing to a mass $\ga 10^6 \msun$ (Begelman 2010; note that the earlier claim in Begelman et al. 2006 that H-burning is unable to postpone collapse is erroneous and is corrected in the later paper). æ

Because of rotation, the black hole that forms initially probably comprises only a small fraction of the core, with a mass $\la 10^3 \msun$, but it grows rapidly at a rate set by the Eddington limit for the massive gaseous envelope. æA novel feature of our model is the prediction that the envelope swells by a factor of $\ga 100$ in radius as it absorbs the energy liberated by black hole growth. æThe resulting object, which we have dubbed a `quasistar', (QS) resembles a red giant with a luminosity comparable to a Seyfert nucleus. æAs the black hole grows inside it, its photosphere expands and cools until it hits a minimum temperature associated with the Hayashi track, at which point it disperses, leaving behind the naked seed black hole.

In this paper, we use a merger-tree approach to estimate the rate at which supermassive stars might form as a function of redshift, and the statistical properties of the resulting QSs and seed black holes. æIn Section 2 we summarize the properties of supermassive stars, QSs and seed black holes, as a function of time and gas infall rate. æThe existence of runaway infall depends on a threshold for global gravitational instability, which we model as a function of the gas mass and angular momentum following each halo merger --- we discuss these criteria in Section 3. æOnce infall is triggered, its rate is determined by the halo potential; the properties of the resulting supermassive star, QS and seed black hole depend on this rate. æ

To relate the properties of the seed black holes to the observable distributions of active and quiescent black holes at different redshifts, we apply a simple set of rules for their subsequent growth by accretion and mergers. These are described in Section 4, and in Section 5 we present our results. æWe compare our direct collapse seed model to a pure Pop III seed model, and also study the case in which both channels of MBH formation operate simultaneously. æWe summarize our conclusions and discuss the prospects of testing this model observationally in Section 6.

\section{Formation of supermassive stars, quasistars and seed black holes}

Theoretical aspects of the formation and evolution of supermassive stars, QSs, and the resulting seed black holes are discussed in Begelman et al. 2006 (BVR2006), \cite{Begelman2008} and \cite{Begelman2010}. æHere we summarize the formulae that we use as inputs for merger-tree calculations presented below, with emphasis on the results from Begelman et al. (2008) and Begelman (2010).

\subsection{Supermassive stars}

A supermassive star in the center of a forming galaxy is unlikely to evolve at constant mass, but rather accumulates mass due to infall. æWe adopt a constant accretion rate 
$\dot M = \dot m \ \msun$ yr$^{-1}$. æFor infall rates larger than about 1 $\msun$ yr$^{-1}$, the star never relaxes thermally to a fully convective state during its ``main sequence" (hydrogen-burning) lifetime; instead it develops a convective core (with polytropic index $n=3$) surrounded by a convectively stable envelope (Begelman 2010). æThe main sequence lifetime is then
\beq
t_{\rm MS} \sim 1.4 \dot m ^{-1/2} T_{c,8}^{-1} \ {\rm Myr},
\label{tms}
\eeq
where $10^8 T_{c,8}$ K is the central core temperature, which depends weakly on the mass and CNO abundance as shown in Fig.~5 of Begelman (2010). æFor the calculations below, we adopt a fiducial value of $T_{c,8}=1.4$, which corresponds to a CNO abundance by mass $Z_{\rm CNO} \sim 10^{-6}$ ($Z/Z_\odot \sim 10^{-4}$). æThen, $t_{\rm MS} \sim \dot m ^{-1/2}$ Myr and the star's mass at the end of hydrogen burning is
\beq
\label{mstar}
M_* \sim 10^6 \dot m^{1/2} \ \msun.
\eeq æææææ

A partially convective supermassive star has a radius $R_* \sim 3.4 \times 10^{13}\dot m$ cm (for electron scattering opacity $\kappa = 0.34$ cm$^2$ g$^{-1}$), independent of mass, and an effective temperature
\beq
T_* = 1.2 \times 10^5 \dot m^{-3/8} \ {\rm K}
\eeq
at the end of hydrogen burning. æAlthough the values of $T_*$ and $R_*$ suggest that supermassive stars may be strong sources of far ultraviolet radiation, we cannot confidently calculate the ionizing flux from a rapidly growing supermassive star because the infalling matter is very optically thick. The opacity is strongly scattering-dominated so there will be a color correction to the effective temperature that hardens the spectrum, but if the infalling matter obscures the supermassive star then the true photosphere could be well outside the star's nominal radius, leading to softer radiation. æBy using a blackbody with temperature $T_*$ and radius $R_*$ to estimate ionizing fluxes, we are effectively assuming that the infalling matter is distributed highly anisotropically in solid angle (e.g., due to rotation).

Because they are strongly dominated by radiation pressure, nonrotating supermassive stars have very small binding energies and can be destabilized by tiny corrections due to general relativity \citep{Hoyle1963,Iben1963a,Iben1963b,Fowler1966}. æHowever, a relatively small amount of rotation (compared to breakup) can increase the binding energy dramatically, stabilizing stars as massive as $\sim 10^8 \msun$ (Fowler 1966). æThe gas falling onto supermassive stars is expected to have substantial, though sub-Keplerian, rotation, and we will conservatively assume that supermassive stars can exist up to masses of $10^7 \msun$, above which they undergo core collapse even if hydrogen is not exhausted.

\subsection{Quasistars and seed black holes}

It is probable that only a tiny fraction of the mass in the core of a supermassive star has low enough angular momentum to collapse directly to a black hole. æAny additional growth of the hole must be accompanied by the outward transport of angular momentum which, barring an efficient relief valve such as a polar jet piercing the star, implies the outward transport of a large amount of energy. The energy liberated by the formation and initial growth of the black hole must be absorbed by the stellar envelope. While the black hole is still small --- perhaps only a few hundred solar masses --- the liberated energy becomes comparable to the binding energy of the envelope, which therefore inflates by a large factor. æIt probably does not disperse, since that would cut off the black hole growth and therefore the energy supply. æ

This leads to a structure dubbed a `quasistar' (Begelman et al. 2006, 2008), which resembles a red giant or Thorne-\.Zytkow object \citep{Thorne1977}
except that the power source for inflating the envelope comes from black-hole accretion instead of nuclear shell burning around a compact core. As the envelope inflates, the core density declines, which reduces the black hole growth rate and the energy feedback rate until a stable equilibrium is reached when the energy output from black hole growth equals the Eddington limit for the total mass of the hole + envelope. æThe effective temperature of this bloated object is 
\beq
\label{QStemp}
T_{\rm QS} \sim 5000 \epsilon_{-1}^{-1/5} m_{\rm BH, 4}^{-2/5} m_{*,6}^{7/20} \ {\rm K},
\eeq æ
where the black hole and QS masses are expressed in units of $10^4 \msun$ and $10^6 \msun$, respectively. The parameter $\epsilon = 0.1 \epsilon_{-1}$ represents an efficiency factor for the conversion of black hole growth into energy. In all of our models below, we adopt $\epsilon_{-1} = 1$; if some of the energy were able to escape without percolating through the envelope, it would imply a smaller $\epsilon$ and allow faster black hole growth during the QS phase. 

Since the black hole grows more rapidly than the QS envelope, $T_{\rm QS}$ decreases with time. æBut it cannot drop below the temperature corresponding to the \cite{Hayashi1961} track. æFor Population III abundances, the minimum temperature is about 4000 K (which we adopt as the fiducial value); for near-solar enrichment, this number drops to $\sim 3000$ K or slightly lower. æUnlike gas pressure-supported red giants, which remain stable as they hover close to the minimum temperature, QSs lose dynamical equilibrium and disperse once they reach this limit. æThis is because QSs are supported against collapse by radiation pressure, which also governs internal energy transport --- implying that there are not enough degrees of freedom available to allow the envelope to adjust quasistatically. æFrom equation (\ref{QStemp}), we see that the black hole mass cannot exceed about 1 percent of the QS mass. æFor $\dot m > 3 \epsilon_{-1}^{8/9} T_{c,8}^{14/9}$, this leads to a final QS mass that substantially exceeds the mass of its supermassive star precursor,
\beq
m_{*,6} \sim 1.4 \epsilon_{-1}^{4/9} \dot m^{8/9}
\eeq
and a black hole mass (at the time of dispersal)
\beq
m_{\rm BH,4} \sim 3 \epsilon_{-1}^{-1/9} \dot m^{7/9}.
\eeq
The lifetime of the QS phase is then 
\beq
t_{\rm QS} \sim 1.4 \epsilon_{-1}^{4/9} \dot m^{-1/9} \ {\rm Myr}.
\eeq
For smaller infall rates, the final QS mass will be similar to the mass of the supermassive star (eq.~[\ref{mstar}]) and the black hole mass will be $\la 10^4 \msun$.

\section{Where quasistars form: modeling the host}
We investigate the formation and evolution of QSs and MBHs via cosmological realizations of the 
merger hierarchy of dark matter halos from early times to the present in a $\Lambda$CDM cosmology \citep[WMAP5,][]{WMAP5}.
We track the merger history of 175 parent halos with present-day masses in the range $10^{11}<M_{\rm h}<10^{15}\,\msun$ æwith a Monte Carlo algorithm \citep{VHM}. The mass resolution of our algorithm reaches $10^5\msun$ at $z=20$, and the most massive halos are split into up to 600,000 progenitors. We also generate 12 Monte Carlo realizations æof the merger hierarchy of a $M_{\rm h}=2\times10^{13}\msun$ halo at $z=6$, starting from $z=25$. 

The formation of QSs requires large gas inflows, $\dot M \ga 1 \msun\,{\rm yr}^{-1}$, within the inner region (sub-parsec scales) of a galaxy. æFollowing BVR2006 we associate these inflows with global dynamical instabilities driven by self-gravity, the so-called `bars within bars' mechanism \citep{Shlosman1989,Shlosman1990}. æSelf-gravitating gas clouds become bar-unstable when the level of rotational support surpasses a certain threshold. A bar can transport angular momentum outward on a dynamical timescale via gravitational and hydrodynamical torques, allowing the radius to shrink. æProvided that the gas is able to cool, this shrinkage leads to even greater instability, on shorter timescales, and the process cascades. æThis mechanism works on a dynamical time ($\approx $ Myr æfor high redshift galaxies) and can operate over many decades of radius (from hundreds of pc to sub-parsec scales). 

Global bar-driven instabilities have now been observed in high-resolution numerical simulations of gas--rich galaxies \citep{Wise2008,Levine2008,Regan2009,Mayer2009}. While BVR2006 suggested that fragmentation and global star formation are suppressed only if the gas temperature remains close to the virial temperature (e.g., due to lack of metals and molecular hydrogen), these simulations find strong inflows that occur before most of the gas fragments and forms stars also at solar metallicities, possibly due to the continuous generation of supersonic turbulence \citep{BegelmanShlosman2009}.
In contrast to BVR2006, we therefore relax the assumption that QSs can be formed only out of zero-metallicity gas. Inspired by \cite{Mayer2009}, we instead assume that inflows are triggered by gas-rich major mergers.

The gas content of galaxies is regulated by the competition between stellar wind/supernova feedback that depletes galaxies of gas, and replenishment of the gas reservoir via mergers with gas-rich halos. In particular, we assume that Pop III stars form in metal-free halos with $T_{\rm vir}> 2000$ K \citep{Yoshida2006}, where we model metal enrichment by the `high feedback, best guess' model of \cite{Scannapieco2003}. 
\cite{Scannapieco2003} model metal enrichment via pair-instability supernovae winds, by following the expansion of spherical outflows into the Hubble flow. They compute the comoving radius, at a given redshift, of an outflow from a population of supernovae that exploded at an earlier time.  Using a modification of the Press--Schechter technique (Scannapieco \& Barkana 2002), they then compute the bivariate mass function of two halos of arbitrary mass and collapse redshift, initially separated by a given comoving distance.  From this function they calculate the number density of supernovae host halos at a given comoving distance from a `repicient' halo of  a given mass $M_h$ that form at a given redshift $z$. By integrating over this function, one can calculate the probability that a halo of mass $M_h$ forms from metal--free gas at a redshift $z$. When a halo forms in our merger tree we calculate the probability that it is metal-free (hence, it can form Pop III stars) and determine if this condition is satisfied using Monte Carlo techniques. If a Pop III star forms in a halo, we conservatively assume that all gas is expelled, leading to a nil gas fraction, $f_{\rm gas}=0$. Gas is replenished only via mergers with gas-rich halos, that is, halos that have never experienced mass loss through Pop III outflows. When two halos merge, we sum their gas masses. We allow QS formation only in mergers where the final gas fraction exceeds a specified threshold, $f_{\rm gas}>f_{\rm thr}$. æWe have varied $f_{\rm thr}$ between 0.025 and 0.25 and found that the model results are largely insensitive to the choice of $f_{\rm thr}$. 

When selecting the sites of QS æformation, we additionally require that the merger remnant has very low angular momentum, which ensures the optimal conditions for gas infall \citep{Begelman2006}. We parameterize the angular momentum of a dark matter halo with mass $M_{\rm h}$ via its halo spin parameter, $\lambda_{\rm spin}\equiv J_{\rm h} E_{\rm h}^{1/2}/ (GM_{\rm h}^{5/2})$, where $J_{\rm h}$ is the total angular momentum and $E_{\rm h}$ is the binding energy. The exact angular momentum threshold below which inflows are triggered, $\lambda_{\rm thr}$, is a free parameter that we constrain against observations (see Section 5 below).

\cite{Donghia2007} investigate the spins of halos that are remnants of major mergers, and compare them to the `global' population using cosmological N-body simulations. They find that halos that are still unrelaxed after a major merger ætend to have higher-than-average spins. They quantify the spin parameter distributions for post-merger halos and relaxed halos. In both cases the distributions are log-normal, but with different parameters: $\overline{\lambda}_{\rm spin}=0.028$ and $\sigma_\lambda=0.58$ for relaxed halos, while æ$\overline{\lambda}_{\rm spin}=0.04$ and $\sigma_\lambda=0.65$ for unrelaxed systems. When a halo forms, we pick its spin parameter from the distribution for relaxed halos. After a major merger, we pick a new $\lambda_{\rm spin}$ from the unrelaxed log-normal distribution, and compare it to $\lambda_{\rm thr}$. If the halo has a spin parameter below the threshold, we consider the halo a candidate for QS formation\footnote{The spin parameter æfor the gaseous component, $\lambda_{\rm gas}$, can be different from the $\lambda$ for the halo. \cite{Gottlober2007} find, however, that statistically $\lambda_{\rm gas}/\lambda_{\rm halo}\simeq 1.4$, and that the tendency is for this ratio to decrease toward unity as redshift increases. We expect, therefore, that statistically our approach is valid.}. 

For halos that meet both criteria, $f_{\rm gas}>f_{\rm thr}$ and $\lambda_{\rm spin}< \lambda_{\rm thr}$, we calculate the gas inflow rate as $\dot{M}= v_{\rm c}^3/G$, where $v_{\rm c}$ is the circular velocity\footnote{A halo of mass $M_{\rm h}$ collapsing at redshift $z$ has a circular velocity \goodbreak
\beq v_{\rm c}= 142 \left[\frac{M_{\rm h}}{10^{12} {\msun} }\right]^{1/3} 
\left[\frac {\Omm}{\Ommz}\ \frac{\Delta_{\rm c}} {18\pi^2}\right]^{1/6} 
(1+z)^{1/2} æ\kms 
\eeq 
where $\Delta_{\rm c}$ is the over-density at virialization relative 
to the critical density. 
For a WMAP5 cosmology we adopt here the æfitting
formula (Bryan \& Norman 1998) $\Delta_{\rm c}=18\pi^2+82 d-39 d^2$, 
where $d\equiv \Ommz-1$ is evaluated at the collapse redshift, so
that $ \Ommz={\Omm (1+z)^3}/[{\Omm (1+z)^3+\Oml+\Omk (1+z)^2}]$. 
}. If $\dot{M} > 1 \msun \ {\rm yr^{-1}}$, then we model supermassive star, QS and seed black hole formation as described in Section 2. We limit the mass of the forming QS to less than the mass of available gas, $f_{\rm gas}M_{\rm h}$.  From stability considerations, we also assume that the supermassive star that precedes the QS undergoes core collapse if its mass exceeds $10^7 \msun$ even if it has not exhausted its core hydrogen.

We further assume that if a seed MBH is already present, QS formation is suppressed. æThis criterion comes from the fact that any QS with a black hole mass exceeding about 1 per cent of the envelope mass will violate the minimum temperature condition and be dispersed by radiation pressure (Begelman et al. 2008). æThus, it is very difficult to lay down a massive envelope of gas around an existing, naked MBH. 

Summarizing, our model assumptions are as follows:
\begin{itemize}
\item inflow triggered by gas-rich major mergers (mass ratio between 1:10 and 1:1);
\item inflow if gas fraction $f_{\rm gas}>f_{\rm thr}$ and spin parameter $\lambda_{\rm spin}< \lambda_{\rm thr}$; 
\item central mass accretion rate $\dot{M}=v_{\rm c}^3/G$;
\item if $\dot{M} > 1 \msun$/yr then QS and seed MBH form, with masses given by
\begin{itemize}
\item $M_{\rm QS}=1.4\times 10 ^6 \epsilon_{-1}^{4/9} \dot m^{8/9}\msun$ 
\item $M_{\rm BH}=3\times 10^4 æ\epsilon_{-1}^{-1/9} \dot m^{7/9}\msun $;
\end{itemize}
\item æQS formation suppressed if MBH already present in halo. 
\end{itemize}

We compare two models, one in which QSs are the only route to MBH seed formation, and one in which Pop III remnants offer a complementary channel. æIn both cases, we assume that Pop III stars form in metal-free halos with $T_{\rm vir}> 2000$ K\footnote{We neglect the effect of a redshift-dependent Lyman--Werner background that progressively increases the minimum mass necessary for Pop III star formation (Trenti \& Stiavelli 2009). As shown below, the differences in the numbers of forming QSs and MBHs, for the cases with and without Pop III remnants, are negligible.  If the Lyman--Werner background reduces the number of forming Pop III stars, the result will fall in-between the two cases considered here.}. In the QS-only scenario, we assume that Pop III stars form but leave no massive remnant \citep[that is, the star mass is below $260\msun$;][]{Fryer2001}. Therefore, Pop III stars in this model only provide stellar wind/supernova feedback that depletes galaxies of gas. In the Pop III remnants+QS scenario we assume a logarithmically flat initial mass function, $dN/d\log M=$const, between $10\msun$ and $600 \msun$, where the upper limit comes from æ\cite{OmukaiPalla}, and æsuppose that seed MBHs form when the progenitor star is in the mass range $40-140 \msun$ or $260-600\msun$ \citep{Fryer2001}. The remnant mass is ætaken to be one-half the mass of the star. æOnce a seed MBH forms, we subsequently treat it in exactly the same way, regardless of its formation mechanism. 

\section{How black hole seeds grow: modeling the cosmic evolution}
We base our model for MBH mass growth on a set of simple assumptions, supported by both simulations of quasar triggering and feedback \citep{Springel2005b}, and analysis of the relationship between MBH masses ($M_{\rm BH}$) and the properties of their hosts in today's galaxies \citep{Gebhardt2000,Ferrarese2000,Ferrarese2002}. The main features of the models have been discussed elsewhere \citep[and references therein]{VHM,VN2009}. æWe summarize in the following the relevant assumptions. æMBHs in galaxies undergoing a major merger (i.e., having a mass ratio $>1:10$) accrete mass and become active. Each MBH accretes an amount of mass, $\Delta M=9\times 10^7(\sigma/200\kms)^{4.24}\msun$, where $\sigma$ is the velocity dispersion after the merger. æThis relationship scales with the $M_{\rm BH}-\sigma$ relation, as it is seen today \citep{Gultekin2009}:
\beq
M_{\rm BH}=1.3\times10^8 æ\left(\frac{\sigma}{200 \kms} \right)^{4.24} \msun;
\eeq
the normalization in $\Delta M$ was chosen to take into account the contribution of mergers, without exceeding the mass given by the $M_{\rm BH}-\sigma$ relation.

We link the correlation between the black hole mass æand the central stellar velocity dispersion of the 
host with the empirical æcorrelation between æthe central stellar velocity dispersion and the asymptotic circular 
velocity ($v_{\rm c}$) of galaxies \citep[Ferrarese 2002; see also][]{Pizzella2005,Baes2003}, 
\beq
\sigma=200 æ\left(\frac{v_{\rm c}}{304 \kms}\right)^{1.19}\kms.
\eeq
The latter is a measure of the total mass of the dark matter halo of the host galaxy. We calculate the circular velocity from the mass of the host halo and its redshift. 

The rate at which mass is accreted scales with the Eddington rate for the MBH, and we set a fixed Eddington ratio of $f_{\rm Edd}=0.3$, based on observations of quasars \citep{Kollmeier2006}. æThis is also the strong upper limit for radiatively efficient spherical accretion onto Pop III remnants \citep{Milos2009}. æAccretion starts after a æhalo dynamical timescale and lasts until the MBH, of initial mass $M_{\rm in}$, has accreted $\Delta M$. æThe lifetime of an AGN therefore depends on how much mass it accretes during each episode:
\begin{equation}
t_{\rm AGN}=\frac{t_{\rm Edd}}{f_{\rm Edd}} \frac{\epsilon}{1-\epsilon}\ln(M_{\rm fin}/M_{\rm in}),
\end{equation}
where $\epsilon$ is the radiative efficiency ($\epsilon \simeq 0.1$), æ$t_{\rm Edd}=0.45$ Gyr and $M_{\rm fin}=\min[(M_{\rm in}+\Delta M),1.3\times10^8 æ(\sigma/200 \kms)^{4.24}\msun]$.

We further assume that, when two galaxies hosting MBHs merge, the MBHs themselves merge within the merger timescale of the host halos, which is a plausible assumption for MBH binaries formed after gas-rich galaxy mergers \citep[and references therein]{Dotti2007}. We adopt the relations suggested by \cite{Taffoni2003} for the merger timescale.

\section{Model results: quasistars and black holes}

\subsection{Constraining the quasistar and black hole formation efficiency}
As described in Section~3, our model depends mostly on one parameter, $\lambda_{\rm thr}$. æWe constrain $\lambda_{\rm thr}$ from below by using the number density of $z\simeq 6$ quasars. æWe obtain a (less certain) upper limit on $\lambda_{\rm thr}$ by estimating the maximum permitted contribution of supermassive stars to reionization, and later show (in Section 5.3) that this constraint is consistent with the So\l tan (1982) constraint on integrated quasar light.

The population of $z\simeq 6$ quasars probes the high-luminosity end of the luminosity function, roughly $L\geqslant 10^{47} {\rm erg\,s^{-1}}$. Recent surveys \citep[and references therein]{Jiang2009} yield a number density of these luminous sources, powered by billion-solar-mass MBHs (assuming sub-Eddington luminosities and negligible beaming), of $\sim$1 Gpc$^{-1}$. This comoving density roughly corresponds to the number density of halos with mass $\simlt10^{13}\msun$ in a WMAP5 cosmology. æWith the scalings described in Section 4, we can express the mass of a halo hosting a $10^9\msun$ MBH as follows:
\beq
M_{\rm h}=5\times10^{13} æ\left[\frac{M_{\rm BH}}{10^9 \msun} \right]^{0.84}
\left[ \frac{\Omm}{\Ommz} \frac{\Delta_{\rm c}} {18\pi^2} \right]^{-1/2} (1+z) ^{-3/2}\msun ,
\eeq
which is in basic agreement with the number density-based estimate. If Pop III remnants, with mass $\sim 100 \msun$, provided the only route to MBH formation, equation~(11) would imply that for an Eddington ratio of 0.3 and a radiative efficiency of 0.1, $t_{\rm AGN}=2.68$ Gyr is needed for a MBH to reach $10^9\msun$. This is more than twice the age of the Universe at $z=6$. Even decreasing $\epsilon$ to 0.05-0.06 (appropriate for either a Schwarzschild BH, or a BH fed by a counter-rotating accretion disc) does not bring the timescale within an acceptable range ($<0.95$ Gyr in a WMAP5 cosmology).
For Pop III remnants to reach a billion solar masses by $z=6$, very specific conditions must be met: accretion occurs {\it continuously} and at rates {\it close to} æ$f_{\rm Edd}=1$, if not higher. If either of these conditions fails, Pop III remnants alone cannot reach the required MBH masses.  Clearly, our approach here is very conservative. The rate at which MBHs in the early Universe accrete is unknown, and it is possible that the most massive high-redshift MBHs may be actually be fed  at the Eddington limit \citep[or above:][]{VR2005}.  One open question is the role of feedback, if accretion is radiatively efficient. \cite{Pelupessy2007,Alvarez2009,Milos2009} suggest that the growth of the MBH is severely limited by thermal feedback to rates not higher than 30\% of the Eddington rate. 

On the other hand, at very high accretion rates, the excess radiation might be effectively trapped, and a black hole accreting at super-critical values does not necessarily radiate at super-Eddington luminosity \citep{Begelman1979,BegelmanMeier1982}.  Also, when the accretion rate is high,  the formation of collimated outflows is common (e.g. blazars). Such collimated jets may not cause feedback directly on the host (which is pierced through), but deposit their kinetic energy at large distances, leaving the host unscathed \citep[in a different context, see][]{Vernaleo2006}. This is likely if at large accretion rates photon trapping decreases the disk luminosity, while concurrently the presence of a jet helps to remove angular momentum, thus promoting efficient accretion (Volonteri, Ghisellini \& Haardt, in preparation).  This is supported by  the strong cosmological evolution of blazars detected by the Burst Alert Telescope (BAT) onboard the {\it Swift} satellite \citep{Ajello2009}.  The derived luminosity function indicates that beamed, jetted sources become dominant at the highest MBH masses ($>10^9 M_\odot$) and redshifts ($z>5$), when compared to the cosmic evolution of radio--quiet sources \citep{Ghisellini2010}. 

\begin{figure}
\includegraphics[width= \columnwidth]{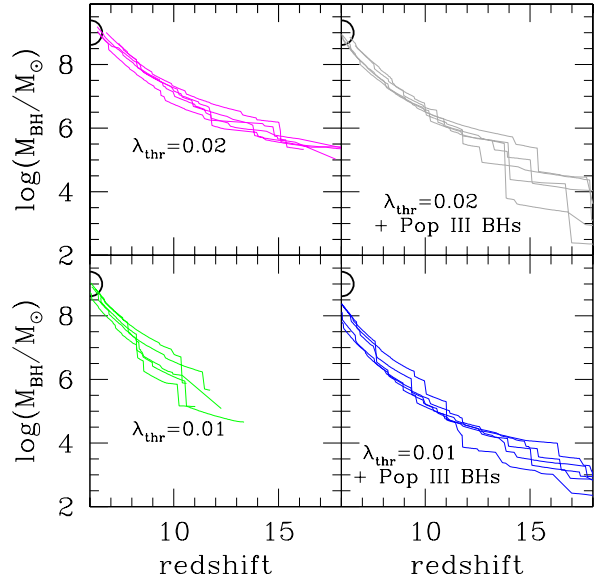}
\caption{Growth of the central black hole in the main halo of merger trees corresponding to $M_{\rm h}=2\times10^{13}\msun$ halos at $z=6$ (we show here only 5 realizations out of 12 for clarity). Lower panels: relatively low-efficiency QS/supermassive star formation case ($\lambda_{\rm thr}=0.01$). Upper panels: efficient QS/supermassive star formation case ($\lambda_{\rm thr}=0.02$). Left panels: Pop III stars leave no remnant that can impede QS and MBH seed formation. Right panels: Pop III stars leave remnant MBHs. The big circle is centered at z=6 and $M_{\rm BH}=10^9\msun$.}
\label{z6}
\end{figure}

Figure~\ref{z6} shows the mass growth of the central black hole in the main halo of merger trees corresponding to $M_{\rm h}=2\times10^{13}\msun$ halos at $z=6$. We see that the requirement set by $z=6$ quasars, powered by $10^9\msun$ MBHs, requires $\lambda_{\rm thr}$ æto be larger than 0.01. 

The case $\lambda_{\rm thr}=0.01$ is only marginally consistent with the constraints. This is because the probability of MBH formation in a given halo is less than $10^{-2}$, given the functional form of the spin parameter probability distribution. A halo needs $\simeq 100$ major mergers to have a probability of MBH formation close to unity (provided, of course, that $v_{\rm c}$ is large enough to provide the required $\dot M$). In the 12 realizations of $M_{\rm h}=2\times10^{13}\msun$ halos at $z=6$, we have not found a single case in which the main halo of the tree forms a MBH if $\lambda_{\rm thr}=0.01$. The MBH hosted in the central halo is instead acquired during a merger with another halo hosting a central MBH. æThis probability is lower if MBH formation is less efficient (lower $\lambda_{\rm thr}$), leading to the main halo acquiring a MBH only at relatively late cosmic times ($z\simeq 12-14$), thus postponing its growth. When Pop III stars leave behind massive remnants, the situation worsens. æSince we assume that QS formation is suppressed in the presence of a MBH, even fewer halos can form massive seeds that can accelerate the growth of a MBH by boosting the merger rate.\footnote{We have here ignored the effect of `gravitational recoils' on the MBH population. For MBHs merging in a gas-rich environment, as expected for these high-redshift systems, \cite{VGD2010} have shown that the probability of ejection is negligible. This is because the accreting gas exerts gravitomagnetic
torques that suffice to align the spins of both MBHs with the angular momentum of the large-scale gas flow, leading to aligned spin-orbit configurations that yield recoil velocities $<100\kms$.  Even for the `worst case scenario' of isotropically distributed spins, the  probability of ejection for MBH binaries in the massive haloes which host very high redshift quasars  effectively drops to zero at $z\simeq 13- 19$.} æ$z=6$ quasars therefore provide a lower limit to QS and seed formation efficiency: $\lambda_{\rm thr}$ æmust be larger than 0.01. 

We estimate the possible contribution of supermassive stars to  hydrogen reionization as follows. æSince the hydrogen recombination timescale  at $z>6$ is shorter than the then-Hubble time, to derive an upper limit to the contribution to reionization we can integrate the number of ionizing photons over time, that is, the sum all the ionizing photons emitted by all the forming supermassive stars. ææTo estimate the number of ionizing photons emitted by a single supermassive star, we assume the photon output of a blackbody with temperature and radius given by $T_*$ and $R_*$ from Section 2, and multiply the output rate by the main sequence lifetime $t_{\rm MS}$ (eq.~[\ref{tms}]).
Since the mass function of QSs --- and the preceding supermassive stars --- peaks at $\simeq 10^6 \msun$ (see Section 5.2 below), we calculate the photon output for $\dot m =1$ and use $T_{c,8} = 1.4$ as before. æThis gives an ionizing photon output of $1.6 \times 10^{67}$ ionizing photons per supermassive star. 

The total number density of ionizing photons emitted by the supermassive star population can then be approximated as:
\begin{equation}
n_{\rm ion}(z) \approx 1.6\times 10^{67} \sum_{i=0,j(z)} n_{{\rm SMS},i} \ æ{\rm ph \ Mpc}^{-3}, 
\end{equation}
where the sum is over the number density of æsupermassive stars ($n_{\rm SMS}$) that form during each timestep between our starting redshift ($z_{\rm max}=20$, timestep 0), and the timestep $j(z)$ corresponding to the redshift of interest, $z$. 

\begin{figure}
\includegraphics[width= \columnwidth]{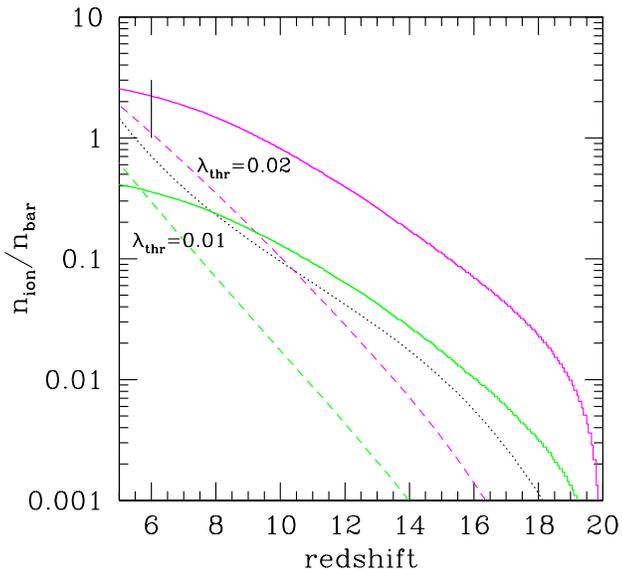}
\caption{Number of reionizing photons per baryon for two QS formation efficiencies, $\lambda_{\rm thr}=0.02$ and $\lambda_{\rm thr}=0.01$. Cases where Pop III remnants offer an additional channel for MBH formation yield almost identical results. Solid curves: supermassive star contribution to the reionizing budget. Dashed curves: contribution from quasars associated with QS remnants where we assume a classic æmulticolor disc spectrum up to $kT_{\rm max}\sim 1\,{\rm keV}\,(M_{\rm BH}/\msun)^{-1/4}$ æ\citep{SS1973}, and a nonthermal power-law component æ$L_\nu\propto \nu^{-\alpha}$, with $\alpha\approx 1$ at higher energies. With this choice, a MBH with mass $\simeq 10^4\msun$ radiates 20\% of its bolometric power as hydrogen-ionizing photons, with $\langle E_\gamma \rangle=140$ eV \citep[for details on the implementation see][]{VG2009}. Dotted curve: contribution from normal stars (from Gnedin 2008).}
\label{reio}
\end{figure}

We compare this number density of photons to the number density of baryons:
\begin{equation}
n_{\rm bar}=f_{\rm bar}\Omm \frac{\rho_{\rm crit}}{\langle m \rangle}\simeq 7.4 \times 10^{66} {\rm Mpc}^{-3},
\end{equation}
where $f_{\rm bar}\simeq 0.17$, $\Omm \simeq 0.3$, æ$\rho_{\rm crit}=9.2\times 10^{-30}$ g cm$^{-3}$, and $\langle m \rangle\simeq 1.7\times 10^{-24}$ g. Therefore,
\begin{equation}
\frac{n_{\rm ion}(z)}{n_{\rm bar}}=\frac{1.6}{0.74}\times \sum_{i=0,j(z)} n_{{\rm SMS},i}.
\end{equation}
This calculation in shown in Figure~\ref{reio}. If we consider a reionization criterion $1<n_{\rm ion}/n_{\rm bar}<3$ at $z=6$ \citep{Gnedin2008}, we see that $\lambda_{\rm thr}>0.02$ tends to overproduce ionizations, provided that the ionizing radiation escapes from the halo core. æAs noted in Section 2, this is far from certain because the gas falling onto the supermassive star is optically thick. æOur estimate of the ionization rate should therefore be regarded as an upper limit, corresponding to the case where the photospheric radiation of the supermassive star is unobscured (e.g., as the result of disc-like inflow) or at least not degraded. æThe ionization rate from supermassive stars therefore provides a weak constraint. æHowever, we note that an additional limit on QS and BH formation, discussed in Section~5.3 (So\l tan's argument), reinforces the limits discussed in this section. æIn the following we assume $0.01<\lambda_{\rm thr}<0.02$, and show results for $\lambda_{\rm thr}=0.01$ (low efficiency) and $\lambda_{\rm thr}=0.02$ (high efficiency).

The quasar population that results from MBHs formed in QSs does not contribute significantly to hydrogen reionization (see Fig.~\ref{reio}) for the low efficiency case, while the high efficiency case produces many more MBHs that produce significant ionizing radiation.  High efficiency MBH formation ($\lambda_{\rm thr}=0.02$) is also constrained to be an upper limit by the contribution of quasars to helium reionization. Taking into consideration ionization energy thresholds for HeII\footnote{The ionization threshold for He I,  24.6 eV, is low enough that He I and H reionization are expected to occur almost concurrently.} and H (54.4 vs.~13.6 eV) and number densities of the two species (12 times more H than He), there are roughly 3 times more He ionizing photons per He atom than photons above 13.6 eV per H atom for our quasar spectrum. As He recombines $\simeq$6 times faster than H, the number of ionizing photons per He atom emitted in one recombination time is only roughly half of the corresponding value for H, and there will be just a small delay between H and He reionizations. 
%So, while the low-efficiency model is consistent with He reionization at $z\simeq 3$, the high-efficiency case indeed represents an upper limit to MBH seed production, as He is reionized at $z\simeq 3.5-5$ (depending on the clumpiness of the intergalactic medium). 
 In the low-efficiency model He reionization occurs at $z\simeq3$, while in the high-efficiency case He is reionized at $z\simeq 3.5-5$ (depending on the clumpiness of the intergalactic medium and the escape fraction of ionizing photons). Although observational constraints on He reionization are still weak, they suggest that HeII reionization occurs at $z\sim 3-4$ \citep[and references therein]{Faucher}. Hence, while the low-efficiency model is consistent with He reionization, the high-efficiency case clearly represents a strong upper limit to MBH seed production, as, if MBHs accrete as efficiently as assumed here, He is reionized at much higher redshift. We note that hard photons from supermassive stars do not contribute much to He reionization (the number of ionizing photons for He atom is roughly one-third of the number of ionizing photons for H atom shown in Fig.~\ref{reio}).

\subsection{Mass functions and formation times of quasistars}
\begin{figure}
\includegraphics[width= \columnwidth]{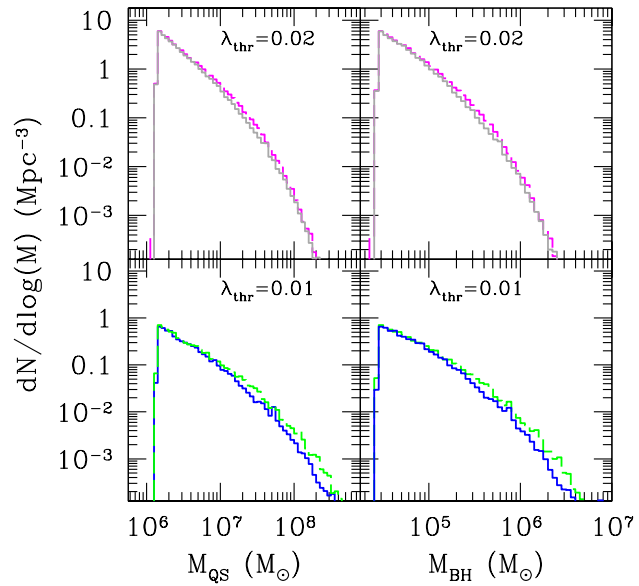}
\caption{Left: mass function of QSs, integrated over all redshifts of formation. Right: mass function of seed black holes. æLower set of histograms: relatively low-efficiency QS formation case ($\lambda_{\rm thr}=0.01$). Upper set of histograms: efficient QS formation case ($\lambda_{\rm thr}=0.02$). Dashed curves: Pop III stars leave no remnant that can impede QS and MBH seed formation. Solid curves: Pop III stars leave remnant MBHs. }
\label{MF_QS}
\end{figure}

Figure~\ref{MF_QS} shows the mass function of QSs, integrated over all redshifts of formation. ææThe mass function peaks at $\simeq 10^6 \msun$, corresponding to the minimum gas infall rate $\dot{M}=v_{\rm c}^3/G = 1 \msun \ {\rm yr^{-1}}$. Since the infall rate is related to the circular velocity of the host halo, it is also linked to the mass of the halo. The halo mass function in a $\Lambda CDM$ cosmology is a steep function of the halo mass; the QS mass function therefore peaks at the lowest allowed mass ($\simeq 10^6 \msun$). 
%The peak at $10^7 \msun$ is due to our assumption that SMSs (hence, QSs) can exist only up to masses of $10^7 \msun$, above which they undergo core collapse.

\begin{figure}
\includegraphics[width= \columnwidth]{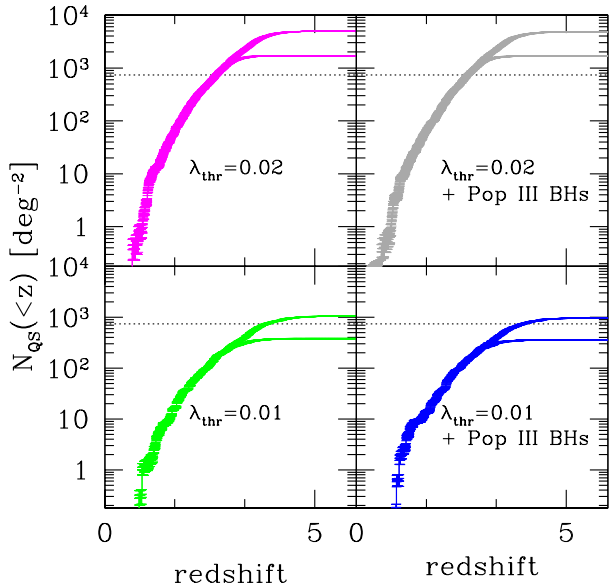}
\caption{Number counts of QSs for a {\it JWST} sensitivity of 10 nJy at 2$\mu$m (upper curves)  and for the  HUDF WFC3/IR in the F160W band, to an AB magnitude limit of 28.8 (Bouwens et al. 2010: lower curves). We assume $T_{\rm QS}=4000$ K. Lower panels: relatively low-efficiency QS formation case ($\lambda_{\rm thr}=0.01$). Upper panels: efficient QS formation case ($\lambda_{\rm thr}=0.02$). Left panels: Pop III stars leave no remnant that can impede QS and MBH seed formation. Right panels: Pop III stars leave remnant MBHs. The dotted horizontal line shows the limit of 1 source per {\it JWST} field ($2.2 \times 2.2$ arcmin$^2$). }
\label{JWST}
\end{figure}

A QS ends its life with a photospheric temperature of æabout 4000 K for Population III abundances (which we adopt as the fiducial value). We model the emission as a blackbody at $T_{\rm QS}=4000$ K for the entire lifetime of the QS (equation 7), and calculate number counts in the 2--10 $\mu$m band of the {\it James Webb Space Telescope} ({\it JWST}), assuming a sensitivity of 10 nJy at 2$\mu$m. These are shown in Figure~\ref{JWST} for our reference models. In the high-efficiency case ($\lambda_{\rm thr}=0.02$), {\it JWST} could detect up to a few QSs per field at $z\simeq 5-10$.  In each panel  of Figure~\ref{JWST}, the lower curves show the expected number counts for fields at the depth of the Hubble Ultra Deep Field, observed with WFC3 on board of the {\it Hubble Space Telecope} in the F160W band, to an AB magnitude limit of 28.8 (Bouwens et al. 2010). We conclude that it is possible that QSs can already be seen in deep WFC3 images, although it is not very likely.

\begin{figure}
\includegraphics[width= \columnwidth]{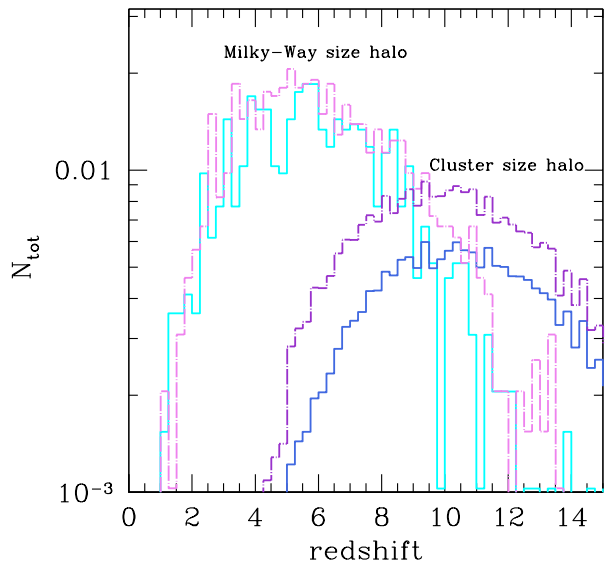}
\caption{Formation times of QSs for two differently biased systems: a MW-size halo and a cluster-size halo. Dot-dashed curves: Pop III stars leave no remnant that can impede QS and MBH seed formation. Solid curves: Pop III stars leave remnant MBHs. Here $N_{\rm tot}$ is normalized to a comoving cubic Mpc. }
\label{bias}
\end{figure}

Although the peak of QS formation occurs at $z>5$, some QSs form at redshift as low as $z\simeq 2$. These late bloomers form in halos with relatively low cosmic bias. We recall that the bias of a halo is the number of standard deviations that the critical collapse overdensity represents on the halo mass scale, $M_{\rm h}$: $\nu_{\rm c}=\delta_{\rm c}(z)/\sigma(M_{\rm h},z)$. æAt any redshift we can identify the characteristic collapsing mass (i.e., $\nu_{\rm c}=1$), and its multiples. The higher $\nu_{\rm c}$, the more massive and rarer the halo, and the higher its bias and clustering strength. Figure~\ref{bias} exemplifies our results for two cases: a Milky Way-sized halo ($M_{\rm h}=2\times 10^{12} \msun$ at $z=0$, lower bias), and a cluster-size halo ($M_{\rm h}=10^{15} \msun$ at $z=0$, higher bias). We therefore expect that the latest-forming QSs should be found in the field, rather than in a high-density environment.  As one goes to lower and lower galaxy masses, the number of halos where QSs can form decreases, as the requirement $\dot M\propto v_c^3>1 \msun/yr$ selects only halos with mass larger than $\sim 10^9 \msun$ as possible QS/MBH hosts. Given that the probability of QS/MBH formation in a given halo is 1-10\% (due to the log-normal distribution of spin parameters), and that major mergers between  $\sim 10^9 \msun$ halos are very rare events in the merger history of galaxies with masses below $\sim 10^{11} \msun$, we expect dwarf galaxies to be rarely seeded with MBHs \citep[results for the frequency of MBHs in dwarf galaxies are very similar to those found in][]{svanwas2010}.  In order to check whether the infrequency of QS/MBH formation at low halo masses is compensated by the larger abundance of small galaxies, we performed 150 merger trees each for halos with mass $5\times 10^9 \msun$, $10^{10}\msun$ and $5\times 10^{10} \msun$. We find that formation of QS/MBHs becomes statistically negligible for halos with mass below  $10^{10}\msun$, and that QSs in low-mass galaxies do not dominate the overall population.

Figure~\ref{MF_QS} also shows the mass function of seed black holes, integrated over all redshifts of formation. æThe mass function peaks at $\simeq$ few æ$10^4\msun$, but rare supermassive seeds, with masses $\simeq 10^6 \msun$, are possible. The more efficient QS and BH formation is, the steeper the mass function. This is because more seeds form at the highest redshifts when the hosts have relatively low mass and low circular velocity.

\subsection{Integrated properties of the black hole population}

\begin{figure}
\includegraphics[width= \columnwidth]{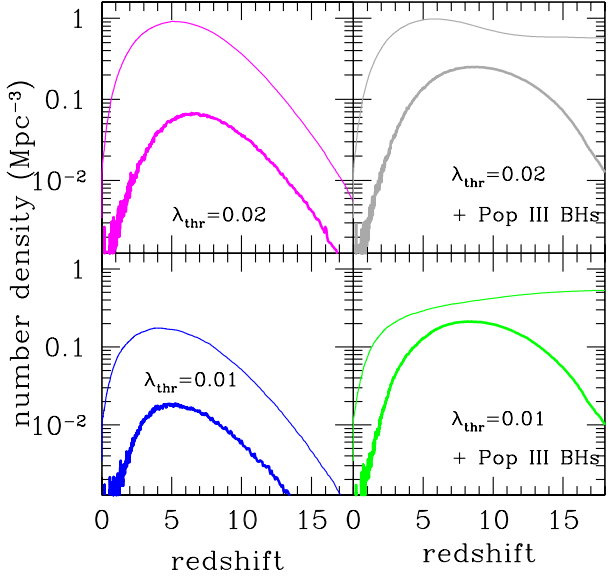}
\caption{Number density of MBHs vs. redshift, integrated over all MBH masses. æLower panel: relatively low-efficiency QS formation case ($\lambda_{\rm thr}=0.01$). Upper panel: efficient QS formation case ($\lambda_{\rm thr}=0.02$). Left panels: Pop III stars leave no remnant that can impede QS and MBH seed formation. Right panels: Pop III stars leave remnant MBHs. Thick curves (the lowest curve in each set): active MBHs only. Note that we do not include information regarding lifetime or a luminosity threshold. Depending on the efficiency of QS and MBH formation ($\lambda_{\rm thr}$), the number densities can be dominated by Pop III or QS MBHs (e.g., QS MBHs always dominate in the $\lambda_{\rm thr}=0.02$ case, while in the $\lambda_{\rm thr}=0.01$ case Pop III remnants can dominate the MBH population {\it in number}.) }
\label{nbh}
\end{figure}

Figure~\ref{nbh} shows the number density of MBHs: depending on the efficiency of QS and MBH formation ($\lambda_{\rm thr}$), the number density can be dominated by either Pop III or QS MBHs. In the low-efficiency case ($\lambda_{\rm thr}=0.01$) the number density of MBHs is dominated by the Pop III channel, if available; the opposite is true for the high-efficiency case. Note how the light Pop III remnant seeds are much more often actively accreting (compare left and right panels). This is an effect of imposing self-regulation of the MBH mass, i.e., that in a given accretion episode a MBH cannot grow to exceed the æ$M_{\rm BH}-\sigma$ relation \citep{VN2009}. Since QS MBHs are born {\it above} the expected $M_{\rm BH}-\sigma$ relation for their host, accretion is effectively damped. Light Pop III remnants, instead, are {\it undermassive} and accrete steadily. 

The number density of MBHs at $z=0$ can be compared to the number density of galaxies, using as a constraint, for instance, the galaxy luminosity function as measured in the Sloan Digital Sky Survey (e.g., Benson et al. 2007). At stellar masses $\sim 10^8 \msun$ the galaxy number density is $\sim 0.3-1 {\rm Mpc}^{-3}$, which is well above our MBH density at $z=0$ (see also Figure 11 in  Carlberg et al. 2009).

\begin{figure}
\includegraphics[width= \columnwidth]{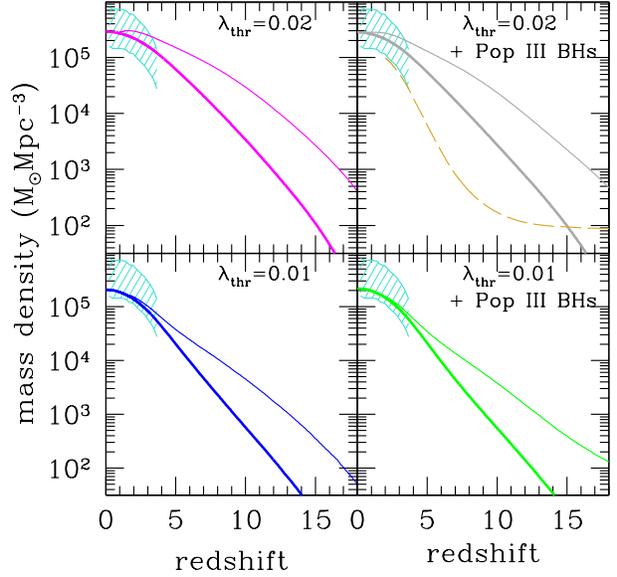}
\caption{Mass density of MBHs vs. redshift. Panels and lines as in Figure~\ref{nbh}. The thick line in each panel shows the integral over time of the accreted mass (cumulative accreted mass density). The dashed curve in the upper-right panel shows the total mass density in MBHs when we consider Pop III remnants only. Hatched area: So\l tan's constraint from the luminosity function of quasars using the Hopkins et al.~(2007) bolometric luminosity function. The boundaries assume a minimum radiative efficiency $\epsilon=0.06$ and a maximum $\epsilon=0.3$ (Schwarzschild to rapidly rotating Kerr MBH). }
\label{rhobh}
\end{figure}

In contrast to the number density of MBHs, the mass density is dominated by QS MBHs (Figure~\ref{rhobh}). We show both the mass density locked into MBHs and the cumulative mass density accreted onto MBHs \citep[related to the So\l tan constraint:][]{Soltan1982}. æSo\l tan's argument offers an additional constraint on our models. We compare the cumulative mass density accreted onto MBHs with that derived from the \cite{Hop_bol_2007} æbolometric luminosity function of quasars:
\begin{equation}
\rho_{\rm QSO}(<z)=\frac {(1-\epsilon)} {\epsilon c^2}
\int_0^{z}\int{{{L' \Phi(L',z)} \over
{H_0(1+z)\sqrt{\Omm(1+z)^3 + \Omega_{\Lambda}}}}}{\rm d}L' {\rm d}z
\end{equation}
where the mass accretion rate,
$\dot M_{\rm acc}=L\epsilon^{-1}c^{-2}$, is converted 
into MBH mass growth,
$\dot M_{\rm BH}=(1-\epsilon)L\epsilon^{-1}c^{-2}$, with $\epsilon$ the energy conversion coefficient. æThe boundaries assume a minimum radiative efficiency $\epsilon=0.06$ and a maximum radiative efficiency $\epsilon=0.3$ (Schwarzschild to rapidly rotating Kerr BH). All our models fit comfortably within the allowed range. We see from Figure~\ref{rhobh} that So\l tan's argument provides an additional constraint to the upper limit on MBH formation via QS seeds. If the QS formation efficiency were much larger than $\lambda_{\rm thr}=0.02$, the cumulative accreted mass density at $z=4$ would be overestimated.

Note that the value of the mass density locked into MBHs at $z=0$, $\rho_{\rm BH,0}=(3.2-5.4)\times 10^5 \msun \ {\rm Mpc}^{-3}$ \citep[e.g.,][and references therein]{Merloni2008} is another, independent constraint that can be evaluated from Figure~\ref{rhobh}. The low-efficiency case is only marginally consistent with $\rho_{\rm BH,0}$. For comparison we show here a case of MBH evolution that considers the Pop III remnant channel only. This is also hardly consistent with $\rho_{\rm BH,0}$ and So\l tan's argument. If BH seeds are rare or very light, a constant $f_{\rm Edd}=0.3$ is insufficient to explain the growth of the BH population and its current demography. 

\begin{figure}
\includegraphics[width= \columnwidth]{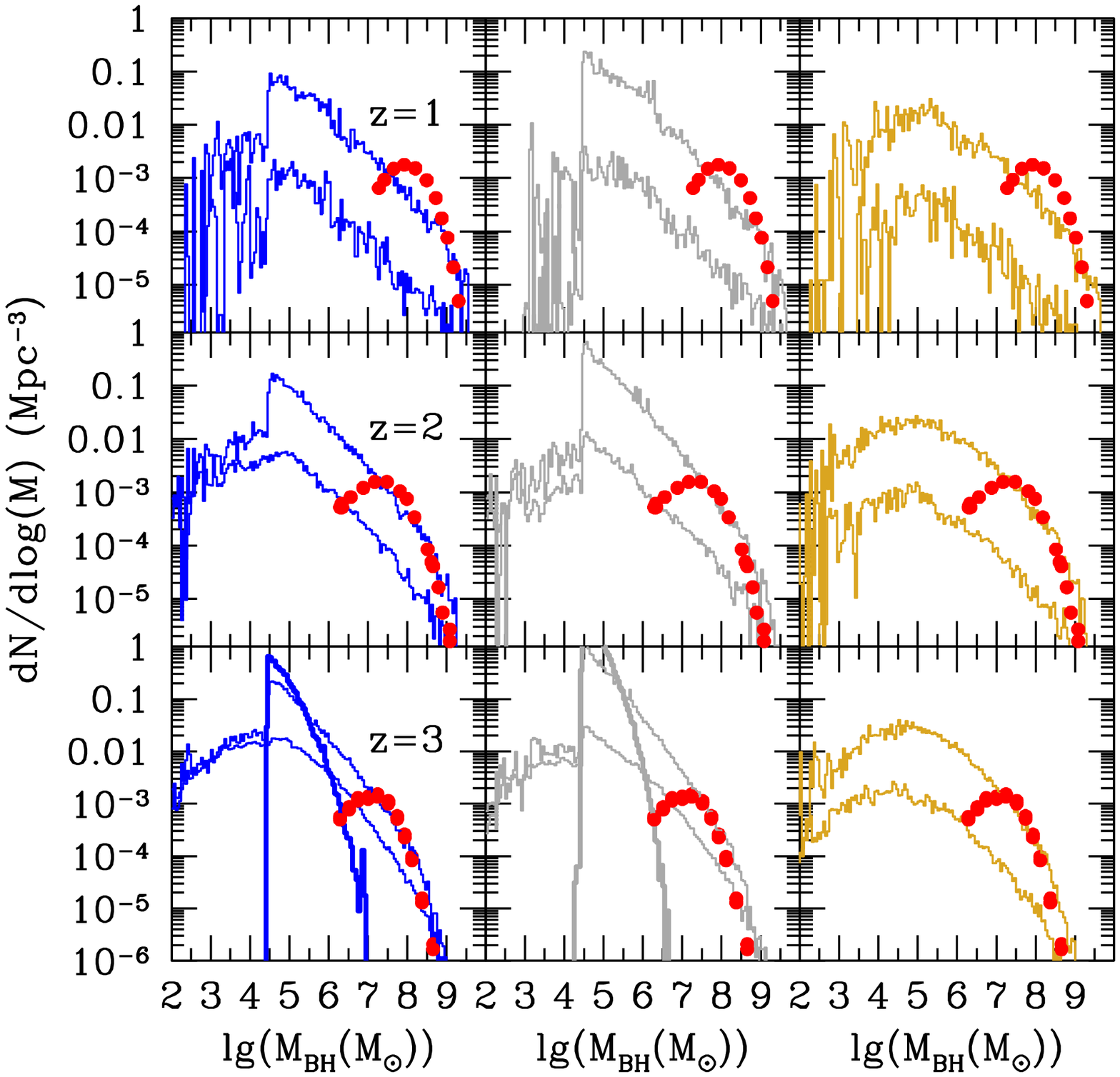}
\caption{ Mass function of MBHs in three redshift bins: $0.95<z<1.025$; $1.9<z<2.1$ and $2.8<z<3.2$. The bin width was chosen to span similar cosmic times ($\sim 0.32$ Gyr).  In each panel we show two curves, the upper curve is the mass function of all black holes. The lower curve is the mass function of active black holes. We also show (big dots) the mass function predicted by Merloni \& Heinz (2008). The very good agreement at masses $>10^7\, \msun$ suggests that our MBH growth prescription is well motivated. Thick lines in the bottom panels: IMF of seed MBHs æleft behind by QSs. Left: $\lambda_{\rm thr}=0.01$ + Pop III MBHs. Middle: $\lambda_{\rm thr}=0.02$ + Pop III MBHs. If we switch off the Pop III channel, the mass function drops to zero at $\log(M_{\rm BH}/\msun)=4.5$, where the seed mass function drops as well. Right: Pop III MBHs only.}
\label{BHMF}
\end{figure}

The mass function of MBHs is shown in Figure~\ref{BHMF}.  We also show the mass function predicted by \cite{Merloni2008}, derived using a completely different technique.  \cite{Merloni2008}  solve the continuity equation for the MBH mass function using the $z=0$ one as a boundary condition, and the  luminosity function as tracer of the MBH growth rate. There is no a priori reason to expect that enforcing the  $M_{\rm BH}-\sigma$ relation and assuming a fixed Eddington rate would match the $ z>0$ mass function of MBHs predicted by  \cite{Merloni2008}. The very good agreement at masses $>10^7\, \msun$ suggests that, albeit very simple, our accretion prescription is well motivated. The disagreement at low masses could be due to incompleteness in the Merloni-Heinz inventory of low-mass MBHs, since their model works backwards from $z=0$, and it does not assume an initial seed mass below which a MBH mass cannot decrease. If the only channel of MBH formation is via QS seeds, then the mass function cuts off below $\log(M_{\rm BH}/\msun)=4.5$, where the seed mass function drops as well. If Pop III remnants offer an alternative route to MBH formation, then the mass function is double-peaked, with each peak æcorresponding to a different mechanism for the formation of seeds. The peak at $\log(M_{\rm BH}/\msun)\simeq4.5$ becomes more pronounced with a higher efficiency of QS (and resulting seed BH) formation. Additional factors, such as mass and redshift dependent accretion rates (we used a simple fixed-Eddington rate model) and obscuration, will modulate the exact shape of the mass function of observable MBHs. 

\section{Discussion and Conclusions}
We have used a merger-tree approach to estimate the rate at which supermassive ($\ga 10^6 \msun$) stars might have formed as a function of redshift, and the statistical properties of the resulting quasistars and seed black holes. æKey to the formation of supermassive stars is a large gas infall rate ($\dot M \simgt 1 \msun$ yr$^{-1}$), driven by global gravitational instability in the potential of a recently merged halo. æWe use the analyses presented by Begelman et al. (2008) and Begelman (2010) to estimate the properties of the resulting supermassive star, QS and seed black hole as a function of this rate. æ

We relate the properties of the seed black holes to the observable distributions of active and quiescent black holes at different redshifts, by applying a simple set of rules for their subsequent growth by accretion and mergers. We compare our direct collapse seed model to a pure Pop III seed model, and also study the case in which both channels for MBH formation operate simultaneously. æ

Our results can be summarized as follows.
\begin{itemize}
%\item The efficiency of supermassive star/QS formation depends mainly on one free parameter, the angular momentum threshold below which inflows are triggered, $\lambda_{\rm thr}$.
\item A limited range of supermassive star/QS/MBH formation efficiencies exists that allows one to reproduce observational constraints (the density of $z=6$ quasars, the cumulative mass density accreted onto MBHs, æthe current mass density of MBHs, reionization). æThese constraints translate into $0.01 \leqslant \lambda_{\rm thr} \leqslant 0.02$.
\item The mass function of QSs peaks at $M_{\rm QS}\simeq 10^6 \msun$, and decreases almost as a power-law with slope $\simeq -(1.5-2)$. The more efficient QS and MBH formation is, the steeper the mass function. This is because more seeds form at the highest redshifts when the hosts have relatively low mass and circular velocity.
\item Modeling QS emission as a blackbody at $T_{\rm QS}=4000$ K, we calculate the number counts for the {\it JWST} $2-10\ \mu$m band. Assuming a sensitivity of 10 nJy at 2$\mu$m, we find that {\it JWST} could detect up to several QSs per field at $z\simeq 5-10$. 
\item The redshift of formation of QSs increases with the cosmic bias of their hosts. We therefore expect that the latest-forming QSs should be found in the field, rather than in a high-density environment. 
\item The number density of MBHs can be dominated by either æthe descendants of Pop III remnants or æQS MBHs. In the low-efficiency case ($\lambda_{\rm thr}=0.01$) the number density of MBHs is dominated by the Pop III channel, if available; the opposite is true for the high-efficiency case ($\lambda_{\rm thr}=0.02$). In contrast to the number density of MBHs, the mass density is always dominated by QS MBHs. 
\item æIf the only channel of MBH formation is via QS seeds, then the mass function of MBHs cuts off below $\log(M_{\rm BH}/\msun)=4.5$, where the seed mass function drops as well. If Pop III remnants offer an alternative route to MBH formation, then the mass function is double-peaked, with each peak tracing æa different seed formation mechanism. æThe peak at $\log(M_{\rm BH}/\msun)\simeq4.5$ becomes more pronounced, the more efficient QS formation is.
\end{itemize}

This model differs from previous proposed mechanisms  (e.g. Lodato \& Natarajan 2006) in various respects. First, gas accumulation in the central regions of protogalaxies is described by global, rather than local, instabilities (e.g., via the Toomre stability criterion). This implies that the spin parameter threshold for collapse cannot be derived by demanding that the Toomre parameter, $Q=\frac{c_{\rm s}\kappa}{\pi G \Sigma}$ (where $\Sigma$ is the  surface mass density, $c_{\rm s}$ is the sound speed,  $\kappa=\sqrt{2}V_{\rm h}/R$ is the epicyclic frequency, and $V$ is the circular velocity of the disc), approaches a critical value, $Q_c$, of order unity. We instead assume a fixed spin parameter threshold (cf. BVR2006). Second,  we have here relaxed the assumption that  large gas infall rates can occur only at zero metallicity to avoid fragmentation. The predicted mass functions have distinctly different shapes (compare Figure 3 with Figure 2 in Lodato \& Natarajan 2007), and MBH seeds tend to form later in the model presented here ($z\simeq 5-10$ instead of $z\simeq 16-18$, when zero metallicity is required). Differences between models can possibly be looked for at the highest redshifts, as at later times the growth of MBHs by accretion and mergers likely washes out dissimilarities. 

The most direct prediction of this work is shown in Figure~\ref{JWST}, where we estimate the detectability of QSs in a {\it JWST} field of view. æAt low metallicities, QSs at redshifts of a few may resemble featureless blackbodies, with colors reminiscent of brown dwarfs. æDirect redshift measurements may not be feasible, in which case quasistars would have to be identified via their massive hosts. æTheir low numbers --- a consequence of their short lifetimes --- will make it even more challenging to find them with a telescope having a relatively small field of view. æIt is possible that some QSs could have formed in metal-enriched regions, and even at relatively low redshifts ($z \ga 2$). æThese might be relatively easy to detect, but extremely rare. Other, secondary characteristics of QSs might conceivably aid in their detection, e.g., if accretion onto the black hole deep in the core leads to the formation of a jet that pierces the stellar surface. æ

The shape of the MBH mass function below $\log(M_{\rm BH}/\msun)=4.5$ is an additional diagnostic. æDirect dynamical measurement of MBH masses at the low-mass end is extremely difficult, due to the necessity of resolving the dynamics of stars within the sphere of influence of the putative MBH, i.e., the region where the Keplerian potential of the MBH dominates over the overall galaxy potential. This region typically subtends much less than an arcsecond in nearby galaxies, æat or below the resolution limit of existing 8--10 m class telescopes. Future 20--30 m telescopes are likely to increase the sample of low-mass MBHs. æ

Gravitational waves produced during the inspirals of compact objects into MBHs --- extreme-mass-ratio inspirals (EMRIs), in particular --- are expected to provide accurate constraints on the mass function of black holes at low redshift. æThe proposed space-based gravitational wave detector, the {\it Laser Interferometer Space Antenna} ({\it LISA}), can probe the mass function in the $10^4M_{\odot}$--$10^7M_{\odot}$ range. \cite{Gair2010} find that with as few as $10$ events, LISA should constrain the slope of the mass function below $\sim 10^6 \msun$ to a precision $\sim0.3$, which is the current level of observational uncertainty in the low-mass slope of the black hole mass function \citep{GreeneHo}. æThe combination of electromagnetic and gravitational wave observations in the coming years will improve the currently limited constraints on what route, or routes, lead to MBH seed formation. 

Numerical simulations \citep[e.g.,][]{Mayer2009} can test how strongly the `bars within bars' inflow mechanism is tied to the angular momentum of the merging galaxies, and either validate our hypothesis that a single parameter $\lambda_{\rm thr}$ can describe the efficiency of the cascade process, or indicate that additional parameters influence the formation of quasistars and their nested black holes. æEven the existence of a second parameter may not have a strong effect on the results: for instance, we found that our results are not very sensitive to varying the threshold gas fraction, as long as it is above $f_{\rm gas}\simeq 0.025$. æIt will also be necessary to study the global dynamical behavior of self-gravitating inflows in the inner regions, where the total gravitational potential is expected to approach a Keplerian shape and the dominant means of angular momentum transport may change from one involving large-scale bars ($m=2$) to a flow resembling an eccentric disk ($m=1$: e.g., Hopkins and Quataert 2009).

It is even more critical to check how much of the infalling gas actually reaches the small radial scales necessary to build the supermassive star. æWe have suggested that fragmentation and star formation may be less important than previously thought in quenching such inflows. æTesting these hypotheses will require numerical experiments with high spatial resolution and a large dynamic range. æ

\section*{Acknowledgments}

MV acknowledges support from the SAO Award TM9-0006X and a Rackham faculty grant. æMCB acknowledges support from NSF grant AST-0907872. MV is grateful to Francesco Haardt for help with reionization models.

%\bibliographystyle{mn2e}
%\bibliography{paper}

\begin{thebibliography}{}

\bibitem[Ajello et al.(2009)]{Ajello2009} Ajello, M., et al.\ 
2009, \apj, 699, 603 

\bibitem[Alvarez et al.(2009)]{Alvarez2009} Alvarez M.~A., Wise 
J.~H., Abel T., 2009, \apjl, 701, L133 

\bibitem[\protect\citeauthoryear{{Baes}, {Buyle}, {Hau} \& {Dejonghe}}{{Baes}
æet~al.}{2003}]{Baes2003}
{Baes} M., æ{Buyle} P., æ{Hau} G.~K.~T., æææ{Dejonghe} H., æ2003, \mnras, 341,
æL44

\bibitem[\protect\citeauthoryear{{Barth}, {Martini}, {Nelson} \& {Ho}}{{Barth}
æet~al.}{2003}]{Barthetal2003}
{Barth} A.~J., æ{Martini} P., æ{Nelson} C.~H., æææ{Ho} L.~C., æ2003, \apjl,
æ594, L95

\bibitem[{{Begelman}(1979)}]{Begelman1979}
{Begelman}, M.~C. 1979, \mnras, 187, 237

\bibitem[{{Begelman} \& {Meier}(1982)}]{BegelmanMeier1982}
{Begelman} M.~C., {Meier} D.~L., 1982, \apj, 253, 873

\bibitem[\protect\citeauthoryear{{Begelman}}{{Begelman}}{2010}]{Begelman2010}
{Begelman} M.~C., æ2010, \mnras, 402, 673

\bibitem[\protect\citeauthoryear{{Begelman}, {Rossi} \& {Armitage}}{{Begelman}
æet~al.}{2008}]{Begelman2008}
{Begelman} M.~C., æ{Rossi} E.~M., æææ{Armitage} P.~J., æ2008, \mnras, 387, 1649

\bibitem[\protect\citeauthoryear{{Begelman} \& {Shlosman}}{{Begelman} \&
æ{Shlosman}}{2009}]{BegelmanShlosman2009}
{Begelman} M.~C., æ{Shlosman} I., æ2009, \apjl, 702, L5

\bibitem[\protect\citeauthoryear{{Begelman}, {Volonteri} \& {Rees}}{{Begelman}
æet~al.}{2006}]{Begelman2006}
{Begelman} M.~C., æ{Volonteri} M ., æææ{Rees} M.~J., æ2006, \mnras, 370, 289

\bibitem[Benson et al.(2007)]{Benson}Benson, A.~J., D{\v z}anovi{\'c}, D., Frenk, C.~S., \& Sharples, R.\ 2007, \mnras, 379, 841 

\bibitem[Bouwens et al.(2010)]{2010ApJ...709L.133B} Bouwens, R.~J., et al.\ 
2010, ApJL, 709, L133 

\bibitem[\protect\citeauthoryear{{Bromm} \& {Loeb}}{{Bromm} \&
æ{Loeb}}{2003}]{BrommLoeb2003}
{Bromm} V., æ{Loeb} A., æ2003, \apj, 596, 34

\bibitem[Bryan 
\& Norman(1998)]{1998ApJ...495...80B} Bryan G.~L., Norman M.~L., 1998, \apj, 495, 80 

\bibitem[Carlberg et al.(2009)]{Carlberg}Carlberg, R. G.,  Sullivan, M., \& Le Borgne, D. 2009, ApJ, 694, 1131

\bibitem[\protect\citeauthoryear{{Djorgovski}, {Volonteri}, {Springel}, {Bromm}
æ\& {Meylan}}{{Djorgovski} et~al.}{2008}]{Djorgovski2008}
{Djorgovski} S.~G., æ{Volonteri} M., æ{Springel} V., æ{Bromm} V., {Meylan}
æG., æ2008, to appear in Proc. XI Marcel Grossmann Meeting on General Relativity, eds. H. Kleinert, R.T. Jantzen \& R. Ruffini, Singapore: World Scientific, arXiv:0803.2862

\bibitem[\protect\citeauthoryear{{D'Onghia} \& {Navarro}}{{D'Onghia} \&
æ{Navarro}}{2007}]{Donghia2007}
{D'Onghia} E., æ{Navarro} J.~F., æ2007, \mnras, 380, L58

\bibitem[\protect\citeauthoryear{{Dotti}, {Colpi}, {Haardt} \& {Mayer}}{{Dotti}
æet~al.}{2007}]{Dotti2007}
{Dotti} M., æ{Colpi} M., æ{Haardt} F., æææ{Mayer} L., æ2007, MNRAS, 379, 956

\bibitem[\protect\citeauthoryear{{Eisenstein} \& {Loeb}}{{Eisenstein} \&
æ{Loeb}}{1995}]{Eisenstein1995}
{Eisenstein} D.~J., æ{Loeb} A., æ1995, \apj, 443, 11

\bibitem[\protect\citeauthoryear{{Fan}}{{Fan}}{2001}]{Fanetal2001a}
{Fan} X. et~al., 2001, AJ, 121, 54


\bibitem[Faucher-Gigu{\`e}re et al.(2008)]{Faucher} 
Faucher-Gigu{\`e}re, C.-A., Prochaska, J.~X., Lidz, A., Hernquist, L., 
\& Zaldarriaga, M.\ 2008, \apj, 681, 831 


\bibitem[\protect\citeauthoryear{{Ferrarese}}{{Ferrarese}}{2002}]{Ferrarese200%
2}
{Ferrarese} L., æ2002, \apj, 578, 90

\bibitem[\protect\citeauthoryear{{Ferrarese} \& {Merritt}}{{Ferrarese} \&
æ{Merritt}}{2000}]{Ferrarese2000}
{Ferrarese} L., æ{Merritt} D., æ2000, \apjl, 539, L9

\bibitem[\protect\citeauthoryear{{Fowler}}{{Fowler}}{1966}]{Fowler1966}
{Fowler} W.~A., æ1966, \apj, 144, 180

\bibitem[\protect\citeauthoryear{{Fryer}, {Woosley} \& {Heger}}{{Fryer}
æet~al.}{2001}]{Fryer2001}
{Fryer} C.~L., æ{Woosley} S.~E., æææ{Heger} A., æ2001, {ApJ}, 550, 372

\bibitem[\protect\citeauthoryear{{Gair}, {Tang} \& {Volonteri}}{{Gair}
æet~al.}{2009}]{Gair2010}
{Gair} J.~R., æ{Tang} C., {Volonteri} M., æ2009, Phys. Rev. D, submitted

\bibitem[\protect\citeauthoryear{{Gebhardt} et~al.,}{{Gebhardt}
æet~al.}{2000}]{Gebhardt2000}
{Gebhardt} K., æet~al., 2000, \apjl, 539, L13

\bibitem[Ghisellini et al.(2010)]{Ghisellini2010} Ghisellini, G., et 
al.\ 2010, \mnras, 442 

\bibitem[\protect\citeauthoryear{{Gnedin}}{{Gnedin}}{2008}]{Gnedin2008}
{Gnedin} N.~Y., æ2008, \apjl, 673, L1

\bibitem[\protect\citeauthoryear{{Gottl{\"o}ber} \& {Yepes}}{{Gottl{\"o}ber} \&
æ{Yepes}}{2007}]{Gottlober2007}
{Gottl{\"o}ber} S., æ{Yepes} G., æ2007, \apj, 664, 117

\bibitem[\protect\citeauthoryear{{Greene} \& {Ho}}{{Greene} \&
æ{Ho}}{2007}]{GreeneHo}
{Greene} J.~E., æ{Ho} L.~C., æ2007, \apj, 667, 131

\bibitem[\protect\citeauthoryear{{G{\"u}ltekin}, {Richstone}, {Gebhardt},
æ{Lauer}, {Tremaine}, {Aller}, {Bender}, {Dressler}, {Faber}, {Filippenko},
æ{Green}, {Ho}, {Kormendy}, {Magorrian}, {Pinkney} \& {Siopis}}{{G{\"u}ltekin}
æet~al.}{2009}]{Gultekin2009}
{G{\"u}ltekin} K., æ{Richstone} D.~O., æ{Gebhardt} K., æ{Lauer} T.~R., {Tremaine} S., æ{Aller} M.~C., æ{Bender} R., æ{Dressler} A., æ{Faber} S.~M., æ{Filippenko} A.~V., æ{Green} R., æ{Ho} L.~C., æ{Kormendy} J., æ{Magorrian} æJ., æ{Pinkney} J., æææ{Siopis} C., æ2009, \apj, 698, 198

\bibitem[\protect\citeauthoryear{{Haehnelt}, {Natarajan} \& {Rees}}{{Haehnelt}
æet~al.}{1998}]{HNR1998}
{Haehnelt} M.~G., æ{Natarajan} P., æææ{Rees} M.~J., æ1998, MNRAS, 300, 817

\bibitem[\protect\citeauthoryear{{Hayashi}}{{Hayashi}}{1961}]{Hayashi1961}
{Hayashi} C., æ1961, \pasj, 13, 450

\bibitem[Hopkins  \& Quataert(2009)]{2009arXiv0912.3257H} Hopkins P.~F., Quataert E., 2009, \mnras, in press, arXiv:0912.3257 

\bibitem[\protect\citeauthoryear{{Hopkins}, {Richards} \&
æ{Hernquist}}{{Hopkins} et~al.}{2007}]{Hop_bol_2007}
{Hopkins} P.~F., æ{Richards} G.~T., æææ{Hernquist} L., æ2007, \apj, 654, 731

\bibitem[\protect\citeauthoryear{{Hoyle} \& {Fowler}}{{Hoyle} \&
æ{Fowler}}{1963}]{Hoyle1963}
{Hoyle} F., æ{Fowler} W.~A., æ1963, \mnras, 125, 169

\bibitem[\protect\citeauthoryear{{Iben} Jr.}{{Iben}}{1963a}]{Iben1963a}
{Iben} Jr. I., æ1963a, \aj, 68, 281

\bibitem[\protect\citeauthoryear{{Iben} Jr.}{{Iben}}{1963b}]{Iben1963b}
{Iben} Jr. I., æ1963b, \apj, 138, 1090

\bibitem[\protect\citeauthoryear{{Jiang}, {Fan}, {Bian}, {Annis}, {Chiu},
æ{Jester}, {Lin}, {Lupton}, {Richards}, {Strauss}, {Malanushenko},
æ{Malanushenko} \& {Schneider}}{{Jiang} et~al.}{2009}]{Jiang2009}
{Jiang} L., æ{Fan} X., æ{Bian} F., æ{Annis} J., æ{Chiu} K., æ{Jester} S.,
æ{Lin} H., æ{Lupton} R.~H., æ{Richards} G.~T., æ{Strauss} M.~A.,
æ{Malanushenko} V., æ{Malanushenko} E., æææ{Schneider} D.~P., æ2009, \aj, 138,
æ305

\bibitem[\protect\citeauthoryear{{Kollmeier}, {Onken}, {Kochanek}, {Gould},
æ{Weinberg}, {Dietrich}, {Cool}, {Dey}, {Eisenstein}, {Jannuzi}, {Le Floc'h}
æ\& {Stern}}{{Kollmeier} et~al.}{2006}]{Kollmeier2006}
{Kollmeier} J.~A., æ{Onken} C.~A., æ{Kochanek} C.~S., æ{Gould} A., æ{Weinberg}
æD.~H., æ{Dietrich} M., æ{Cool} R., æ{Dey} A., æ{Eisenstein} D.~J., æ{Jannuzi}
æB.~T., æ{Le Floc'h} E., æææ{Stern} D., æ2006, \apj, 648, 128

\bibitem[\protect\citeauthoryear{{Komatsu}, {Dunkley}, {Nolta}, {Bennett},
æ{Gold}, {Hinshaw}, {Jarosik}, {Larson}, {Limon}, {Page}, {Spergel},
æ{Halpern}, {Hill}, {Kogut}, {Meyer}, {Tucker}, {Weiland}, {Wollack} \&
æ{Wright}}{{Komatsu} et~al.}{2009}]{WMAP5}
{Komatsu} E., æ{Dunkley} J., æ{Nolta} M.~R., æ{Bennett} C.~L., æ{Gold} B.,
æ{Hinshaw} G., æ{Jarosik} N., æ{Larson} D., æ{Limon} M., æ{Page} L.,
æ{Spergel} D.~N., æ{Halpern} M., æ{Hill} R.~S., æ{Kogut} A., æ{Meyer} S.~S.,
æ{Tucker} G.~S., æ{Weiland} J.~L., æ{Wollack} E., æææ{Wright} E.~L., æ2009,
æ\apjs, 180, 330

\bibitem[\protect\citeauthoryear{{Koushiappas}, {Bullock} \&
æ{Dekel}}{{Koushiappas} et~al.}{2004}]{Koushiappas2004}
{Koushiappas} S.~M., æ{Bullock} J.~S., æææ{Dekel} A., æ2004, \mnras, 354, 292

\bibitem[\protect\citeauthoryear{{Lagos}, {Cora} \& {Padilla}}{{Lagos}
æet~al.}{2008}]{Lagos2008}
{Lagos} C.~D.~P., æ{Cora} S.~A., æææ{Padilla} N.~D., æ2008, \mnras, 388, 587

\bibitem[\protect\citeauthoryear{{Levine}, {Gnedin}, {Hamilton} \&
æ{Kravtsov}}{{Levine} et~al.}{2008}]{Levine2008}
{Levine} R., æ{Gnedin} N.~Y., æ{Hamilton} A.~J.~S., æææ{Kravtsov} A.~V., æ2008,
æ\apj, 678, 154

\bibitem[\protect\citeauthoryear{{Lodato} \& {Natarajan}}{{Lodato} \&
æ{Natarajan}}{2006}]{LN2006}
{Lodato} G., æ{Natarajan} P., æ2006, \mnras, 371, 1813

\bibitem[Lodato 
\& Natarajan(2007)]{2007MNRAS.377L..64L} Lodato G., Natarajan P., 2007, \mnras, 377, L64 

\bibitem[\protect\citeauthoryear{{Loeb} \& {Rasio}}{{Loeb} \&
æ{Rasio}}{1994}]{LoebRasio1994}
{Loeb} A., æ{Rasio} F.~A., æ1994, {\apj}, 432, 52

\bibitem[\protect\citeauthoryear{{Madau} \& {Rees}}{{Madau} \&
æ{Rees}}{2001}]{MadauRees2001}
{Madau} P., æ{Rees} M.~J., æ2001, \apjl, 551, L27

\bibitem[\protect\citeauthoryear{{Mayer}, {Kazantzidis}, {Escala} \&
æ{Callegari}}{{Mayer} et~al.}{2009}]{Mayer2009}
{Mayer} L., æ{Kazantzidis} S., æ{Escala} A., æææ{Callegari} S., æ2009, Nature, in press, arXiv:0912.4262 

\bibitem[\protect\citeauthoryear{{Merloni} \& {Heinz}}{{Merloni} \&
æ{Heinz}}{2008}]{Merloni2008}
{Merloni} A., æ{Heinz} S., æ2008, \mnras, 388, 1011

\bibitem[\protect\citeauthoryear{{Milosavljevi{\'c}}, {Couch} \&
æ{Bromm}}{{Milosavljevi{\'c}} et~al.}{2009}]{Milos2009}
{Milosavljevi{\'c}} M., æ{Couch} S.~M., æææ{Bromm} V., æ2009, \apjl, 696, L146

\bibitem[\protect\citeauthoryear{{Monaco}, {Fontanot} \& {Taffoni}}{{Monaco}
æet~al.}{2007}]{Monaco2007}
{Monaco} P., æ{Fontanot} F., æææ{Taffoni} G., æ2007, \mnras, 375, 1189

\bibitem[\protect\citeauthoryear{{Omukai} \& {Palla}}{{Omukai} \&
æ{Palla}}{2003}]{OmukaiPalla}
{Omukai} K., æ{Palla} F., æ2003, \apj, 589, 677

\bibitem[Pelupessy et al.(2007)]{Pelupessy2007} Pelupessy F.~I., Di 
Matteo T., Ciardi B., 2007, \apj, 665, 107 

\bibitem[\protect\citeauthoryear{{Pizzella}, {Corsini}, {Dalla Bont{\`a}},
æ{Sarzi}, {Coccato} \& {Bertola}}{{Pizzella} et~al.}{2005}]{Pizzella2005}
{Pizzella} A., æ{Corsini} E.~M., æ{Dalla Bont{\`a}} E., æ{Sarzi} M., æ{Coccato}
æL., æææ{Bertola} F., æ2005, \apj, 631, 785

\bibitem[\protect\citeauthoryear{{Rees}}{{Rees}}{1978}]{Rees1978}
{Rees} M.~J., æ1978, æin Structure
æand Properties of Nearby Galaxies IAU Symp.~77, eds.~{E.~M.~Berkhuijsen \& R.~Wielebinski}, Dordrecht: Reidel, 237 

\bibitem[\protect\citeauthoryear{{Regan} \& {Haehnelt}}{{Regan} \&
æ{Haehnelt}}{2009}]{Regan2009}
{Regan} J.~A., æ{Haehnelt} M.~G., æ2009, \mnras, 393, 858

\bibitem[\protect\citeauthoryear{{Rhook} \& {Haehnelt}}{{Rhook} \&
æ{Haehnelt}}{2006}]{Rhook2006}
{Rhook} K.~J., æ{Haehnelt} M.~G., æ2006, \mnras, 373, 623

\bibitem[\protect\citeauthoryear{{Scannapieco} \&
æ{Barkana}}{{Scannapieco} et~al.}{2002}]{Scannapieco2002}Scannapieco E.,  Barkana R.,  2002, ApJ, 571, 585 

\bibitem[\protect\citeauthoryear{{Scannapieco}, {Schneider} \&
æ{Ferrara}}{{Scannapieco} et~al.}{2003}]{Scannapieco2003}
{Scannapieco} E., æ{Schneider} R., æææ{Ferrara} A., æ2003, ApJ, 589, 35

\bibitem[\protect\citeauthoryear{{Shakura} \& {Sunyaev}}{{Shakura} \& {Sunyaev}}{1973}]{SS1973} {Shakura} N.~I., æ{Sunyaev} R.~A., æ1973, \aap, 24, 337

\bibitem[\protect\citeauthoryear{{Shapiro}}{{Shapiro}}{2005}]{Shapiro2005}
{Shapiro} S.~L., æ2005, \apj, 620, 59

\bibitem[\protect\citeauthoryear{{Shlosman}, {Begelman} \& {Frank}}{{Shlosman}
æet~al.}{1990}]{Shlosman1990}
{Shlosman} I., æ{Begelman} M.~C., æææ{Frank} J., æ1990, \nat, 345, 679

\bibitem[\protect\citeauthoryear{{Shlosman}, {Frank} \& {Begelman}}{{Shlosman}
æet~al.}{1989}]{Shlosman1989}
{Shlosman} I., æ{Frank} J., æææ{Begelman} M.~C., æ1989, \nat, 338, 45

\bibitem[\protect\citeauthoryear{{So{\l}tan}}{{So{\l}tan}}{1982}]{Soltan1982}
{So{\l}tan} A., æ1982, \mnras, 200, 115

\bibitem[\protect\citeauthoryear{{Somerville}, {Hopkins}, {Cox}, {Robertson} \&
æ{Hernquist}}{{Somerville} et~al.}{2008}]{Somerville2008}
{Somerville} R.~S., æ{Hopkins} P.~F., æ{Cox} T.~J., æ{Robertson} B.~E.,
æ{Hernquist} L., æ2008, \mnras, 391, 481

\bibitem[\protect\citeauthoryear{{Springel}, {Di Matteo} \&
æ{Hernquist}}{{Springel} et~al.}{2005}]{Springel2005b}
{Springel} V., æ{Di Matteo} T., æææ{Hernquist} L., æ2005, \mnras, 361, 776

\bibitem[\protect\citeauthoryear{{Taffoni}, {Mayer}, {Colpi} \&
æ{Governato}}{{Taffoni} et~al.}{2003}]{Taffoni2003}
{Taffoni} G., æ{Mayer} L., æ{Colpi} M., æææ{Governato} F., æ2003, MNRAS, 341,
æ434

\bibitem[\protect\citeauthoryear{{Thorne} \& {\.Zytkow}}{{Thorne} \&
æ{\.Zytkow}}{1977}]{Thorne1977}
{Thorne} K.~S., æ{\.Zytkow} A.~N., æ1977, \apj, 212, 832

\bibitem[\protect\citeauthoryear{{Trenti} \& {Stiavelli}}{{Trenti} \&
æ{Stiavelli}}{2009}]{Trenti2009} Trenti \& Stiavelli, 2009, ApJ 694, 879 

\bibitem[\protect\citeauthoryear{{Van Wassenhove}, {Volonteri}, {Walker} \&
æ{Gair}}{{Van Wassenhove} et~al.}{2010}]{svanwas2010}
{Van Wassenhove} S., æ{Volonteri} M., æ{Walker} M.~G., {Gair} J.~R., æ2010, \mnras, in press, æarXiv:1001.5451

\bibitem[Vernaleo 
\& Reynolds(2006)]{Vernaleo2006} Vernaleo J.~C., Reynolds C.~S., 2006, \apj, 645, 83 

\bibitem[\protect\citeauthoryear{{Volonteri} \& {Gnedin}}{{Volonteri} \&
æ{Gnedin}}{2009}]{VG2009}
{Volonteri} M., æ{Gnedin} N.~Y., æ2009, \apj, 703, 2113

\bibitem[\protect\citeauthoryear{{Volonteri}, {Gultekin} \&
æ{Dotti}}{{Volonteri} et~al.}{2010}]{VGD2010}
{Volonteri} M., æ{Gultekin} K., æææ{Dotti} M., æ2010, \mnras, 404, 2143

\bibitem[\protect\citeauthoryear{{Volonteri}, {Haardt} \& {Madau}}{{Volonteri}
æet~al.}{2003}]{VHM}
{Volonteri} M., æ{Haardt} F., æææ{Madau} P., æ2003, \apj, 582, 559

\bibitem[\protect\citeauthoryear{{Volonteri}, {Lodato} \&
æ{Natarajan}}{{Volonteri} et~al.}{2008}]{VLN2008}
{Volonteri} M., æ{Lodato} G., æææ{Natarajan} P., æ2008, \mnras, 383, 1079

\bibitem[\protect\citeauthoryear{{Volonteri} \& {Natarajan}}{{Volonteri} \&
æ{Natarajan}}{2009}]{VN2009}
{Volonteri} M., æ{Natarajan} P., æ2009, \mnras, 400, 1911

\bibitem[\protect\citeauthoryear{{Volonteri} \& {Rees}}{{Volonteri} \&
æ{Rees}}{2005}]{VR2005}
{Volonteri} M., æ{Rees} M.~J., æ2005, \apj, 633, 624

\bibitem[\protect\citeauthoryear{{Volonteri} \& {Rees}}{{Volonteri} \&
æ{Rees}}{2006}]{VR2006}
{Volonteri} M., æ{Rees} M.~J., æ2006, \apj, 650, 669

\bibitem[\protect\citeauthoryear{{Willott}, {Delorme}, {Reyl{\'e}}, {Albert},
æ{Bergeron}, {Crampton}, {Delfosse}, {Forveille}, {Hutchings}, {McLure},
æ{Omont} \& {Schade}}{{Willott} et~al.}{2009}]{Willott2009}
{Willott} C.~J., æ{Delorme} P., æ{Reyl{\'e}} C., æ{Albert} L., æ{Bergeron} J.,
æ{Crampton} D., æ{Delfosse} X., æ{Forveille} T., æ{Hutchings} J.~B., æ{McLure}
æR.~J., æ{Omont} A., æææ{Schade} D., æ2009, \aj, 137, 3541

\bibitem[\protect\citeauthoryear{{Wise}, {Turk} \& {Abel}}{{Wise}
æet~al.}{2008}]{Wise2008}
{Wise} J.~H., æ{Turk} M.~J., æææ{Abel} T., æ2008, \apj, 682, 745

\bibitem[\protect\citeauthoryear{{Yoshida}, {Omukai}, {Hernquist} \&
æ{Abel}}{{Yoshida} et~al.}{2006}]{Yoshida2006}
{Yoshida} N., æ{Omukai} K., æ{Hernquist} L., æææ{Abel} T., æ2006, \apj, 652, 6

\end{thebibliography}

\bsp

\label{lastpage}

\end{document}